# Sensitive and broadband measurement of dispersion in a cavity using a Fourier transform spectrometer with kHz resolution

LUCILE RUTKOWSKI,[1] ALEXANDRA C. JOHANSSON,[1] GANG ZHAO,[1,2] THOMAS HAUSMANINGER,[1] AMIR KHODABAKHSH,[1] OVE AXNER,[1] AND ALEKSANDRA FOLTYNOWICZ[1,*]

[1]Department of Physics, Umeå University, 901 87 Umeå, Sweden
[2] State Key Laboratory of Quantum Optics and Optics Devices, Institute of Laser Spectroscopy, Shanxi University, Taiyuan 030006, China
*Corresponding author: aleksandra.foltynowicz@umu.se

**Optical cavities provide high sensitivity to dispersion since their resonance frequencies depend on the index of refraction. We present a direct, broadband, and accurate measurement of the modes of a high finesse cavity using an optical frequency comb and a mechanical Fourier transform spectrometer with a kHz-level resolution. We characterize 16000 cavity modes spanning 16 THz of bandwidth in terms of center frequency, linewidth, and amplitude. We retrieve the group delay dispersion of the cavity mirror coatings and pure $N_2$ with 0.1 fs$^2$ precision and 1 fs$^2$ accuracy, as well as the refractivity of the $3\nu_1+\nu_3$ absorption band of $CO_2$ with $5 \times 10^{-12}$ precision. This opens up for broadband refractive index metrology and calibration-free spectroscopy of entire molecular bands.**

Fabry-Perot cavities in combination with narrow linewidth continuous wave (cw) lasers are versatile tools for ultra-sensitive measurements of displacement, absorption, and dispersion. For example, high precision measurements of minute length variation of Fabry-Perot cavities enable detection of gravitational waves [1]. A pressure sensor based on the measurement of gas refractivity inside a cavity can outperform a manometer [2]. Cavity-enhanced molecular absorption [3, 4] and dispersion [5, 6] spectroscopies, which rely on the measurement of intracavity absorption losses and dispersion induced shifts of the cavity modes, respectively, provide complementary information about the molecular transitions and high sensitivity to absorption/dispersion. However, cw lasers allow such measurements only over narrow bandwidths, typically in the sub-THz range. Optical frequency combs, whose spectra consist of thousands of equidistant narrow lines, can probe cavity modes over a much broader bandwidth. In cavity-enhanced optical frequency comb absorption spectroscopy, spectra of entire molecular bands can be acquired with high resolution in short acquisition times [7-10]. Combs are also an ideal tool for measurements of broadband cavity dispersion induced either by the cavity mirror coatings or intracavity samples. However, previous demonstrations [11-13] did not fully benefit from the high frequency accuracy provided by the comb and suffered from poor spectral resolution (at the THz level).

Recent advances in comb-based Fourier transform spectroscopy have provided means to measure spectra over the entire comb bandwidth with resolution directly given by the comb linewidth, using either dual-comb spectrometers [14-17] or mechanical Fourier transform spectrometers (FTS) [18]. Here we use a frequency comb and a mechanical FTS with sub-nominal resolution [18] to directly measure broadband transmission spectra of a high finesse cavity with high signal-to-noise ratio and frequency precision and accuracy. We fully characterize the cavity modes in terms of amplitude, width, and center frequency. From the shift of the cavity mode frequencies we determine the group delay dispersion of the cavity mirror coatings and intracavity gas samples, including the resonant dispersion induced by molecular transitions.

The cavity mode frequencies, $\nu_q$, fulfill the resonance condition on the round-trip phase shift of the electric field, which can be written as

$$\Phi(\nu_q) = \phi_0(\nu_q) + \phi_n(\nu_q) = 2\pi q, \qquad (1)$$

where q is the integer mode index, $\phi_0(\nu_q) = 2\pi\nu_q 2L/c + \phi_m$ is the phase shift inside an empty cavity, where L is the cavity length, c is the speed of light, and $\phi_m$ is the phase shift induced by the cavity mirror coatings, and $\phi_n(\nu_q) = 2\pi\nu_q [n(\nu_q)-1] 2L/c$ is the phase shift induced by the intracavity sample with refractive index n. The cavity modes are not evenly spaced because $\Phi$ varies non-linearly with frequency. To evaluate the variation of the cavity mode spacing, we define a reference cavity mode scale as

$$\nu_q^0 = q FSR_{ref}^0 + f_0, \qquad (2)$$

where $FSR_{ref}^0 = 2\pi [(\partial \phi_0/\partial \nu)_{ref}]^{-1}$ is the empty cavity free spectral range (FSR) at a reference frequency $\nu_{ref}$, $q = \text{floor}[\nu_q/FSR_{ref}^0]$, and $f_0 = \nu_{ref} - q_{ref} FSR_{ref}^0$ is the empty cavity offset frequency defined at $\nu_{ref}$ so that $\nu_{ref}^0 = \nu_{ref}$. The relation between the shift of the cavity modes, $\Delta\nu = \nu_q - \nu_q^0$, and the cavity phase shift, Eq. (1), can be found by Taylor expanding $\Phi$ around $\nu_{ref}$ to first order, which yields

$$\Delta \nu = \nu_q - \nu_q^0 = \frac{FSR_{ref}}{2\pi}\left[2\pi q - \Phi(\nu_q^0)\right], \quad (3)$$

where $FSR_{ref} = 2\pi [(\partial \Phi/\partial \nu)_{ref}]^{-1}$. The intracavity group delay dispersion (GDD) is then calculated as

$$GDD(\nu_q^0) \equiv \frac{1}{4\pi^2}\frac{\partial^2 \Phi}{\partial \nu^2} = -\frac{1}{2\pi FSR_{ref}}\frac{\partial^2 \Delta\nu}{\partial \nu^2}. \quad (4)$$

The mode shift in an empty cavity, $\Delta\nu_0$, originates from the dispersion of the cavity mirror coatings. When the cavity is filled with a gas, an additional shift, $\Delta\nu_n$, occurs because of the frequency dependence of the refractive index

$$\Delta\nu_n = -\nu_q^0\left[n(\nu_q^0) - 1\right]. \quad (5)$$

The refractive index n has two contributions $n = n_{na} + n_{abs}$, where $n_{na}$ is the slowly varying refractive index of a non-absorbing gas, given by the Sellmeier equation [19], while $n_{abs}$ is the refractive index of molecular transitions, defined as

$$n_{abs}(\nu_q^0) - 1 = -\frac{c}{4\pi\nu_q^0}\rho \sum_i S_i \, \text{Im}\left[\chi_i(\nu_q^0)\right], \quad (6)$$

where $\rho$ is the absorbing sample density, $S_i$ is the line intensity and $\chi_i$ is the complex line shape function of the $i^{th}$ transition.

The experimental setup is depicted in Fig. 1(a). The cavity with a finesse of ~2000 consisted of two dielectric mirrors with 5 m radius of curvature separated by L = 45 cm, yielding an FSR of 333 MHz. The mirrors were glued to two ends of a stainless steel tube, connected to a gas system, and a ring piezoelectric-transducer (PZT) was inserted between one of the mirrors and the tube to control the cavity length. The cavity modes were probed by an amplified Er:fiber frequency comb with a repetition rate ($f_{rep}$) of 250 MHz. Because of the difference between the cavity FSR and comb $f_{rep}$ the cavity acted as a filter for the comb, as shown in Fig. 1(b), and the repetition rate in cavity transmission was equal to $f_{rep}^T = 4 f_{rep} = 3 FSR = 1$ GHz. The comb offset frequency, $f_{ceo}$, was stabilized by locking the output of an f-2f interferometer to a frequency provided by a GPS-referenced Rubidium clock, $f_{clk} = 20$ MHz, via feedback to the current of a diode laser pumping the Er:fiber oscillator. The frequency of one of the comb lines was stabilized to a narrow-linewidth cw Er:fiber laser locked to the $P_e(8)$ $CO_2$ transition at $\lambda_{cw} = 1576.9396$ nm using sub-Doppler noise-immune cavity-enhanced optical heterodyne molecular spectroscopy (NICE-OHMS [20], not shown in the figure). The linewidth of the cw laser was 2.8(6) kHz, estimated using the method described in [21]. The two laser beams were combined in a fiber and the beat note between the cw laser and the closest comb line was detected in free space after dispersing the spectrum with a grating. The beatnote was locked to a tunable RF frequency, $f_{DDS}$, generated by a direct digital synthesizer (DDS) referenced to the Rb clock, via feedback to an intracavity PZT and an electro-optic modulator (EOM) controlling the $f_{rep}$. This optical lock transferred the linewidth of the cw laser to the closest comb line, which was verified by measuring the width of the optical beat note below the Hz level using a spectrum analyzer. The cavity length was stabilized by locking one of the resonances to the cw laser using the Pound-Drever-Hall (PDH) technique. This involved phase modulation of the electric field of the cw laser using an EOM at a frequency $f_{PDH} = 20$ MHz and phase sensitive detection of the cavity reflected light, which was dispersed with a grating to avoid detector saturation by the reflected comb intensity. The correction signal was fed to the cavity PZT with a closed-loop bandwidth of 1.1 kHz. Due to this relatively low bandwidth a frequency jitter of the cavity modes remained, yielding a mode broadening of 20 kHz, calculated from the power spectral density of the closed-loop error signal [21].

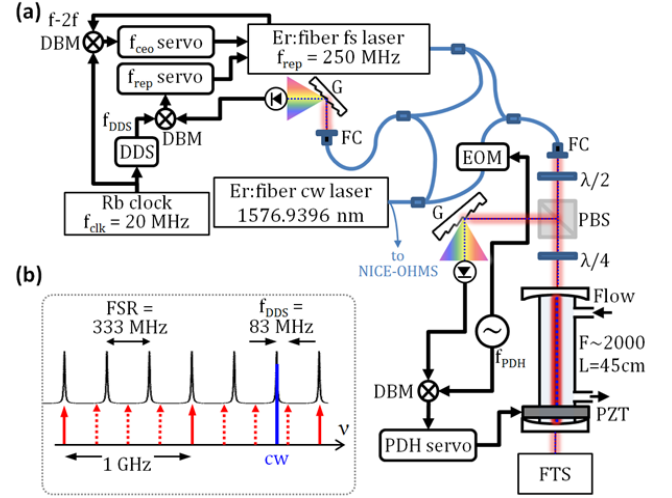

Fig. 1. (a) Experimental setup: f-2f, f-2f interferometer beat note; DBM, double-balanced mixer; DDS, direct digital synthesizer; FC, fiber collimator; λ/2, half-waveplate; PBS, polarizing beam splitter; λ/4, quarter-waveplate; PZT, piezoelectric transducer; FTS, Fourier transform spectrometer; EOM, electro-optic modulator; PDH, Pound-Drever-Hall G, grating; (b) Matching of the comb (red) and the cw laser (blue) to the cavity (black). The reflected comb lines are shown with the red dashed lines.

Setting $f_{DDS}$ close to 83 MHz, as shown in Fig. 1(b), brought the comb lines close to resonance with the cavity modes. The transmitted light, containing both the comb and the cw laser frequencies, was analyzed with a fast-scanning mechanical FTS equipped with an auto-balancing detector [8, 22]. The nominal resolution of the FTS was set to $f_{rep}^T/3 = 333$ MHz, which allowed precise sampling of the intensities of the individual comb lines and the cw laser without distortion induced by the instrumental line shape of the FTS [18]. For a given $f_{rep}$ value, the spectrum contained 1 spectral element per transmitted comb line, separated by 2 sampling points with intensities at the noise level (except for the point sampling the cw laser frequency). The spectrum of the cavity modes was obtained by interleaving spectra recorded with different $f_{rep}$ values. The $f_{rep}$ was stepped by tuning $f_{DDS}$ with a step of 20 kHz, which yielded more than 10 points per cavity mode width. Note that this step is not the resolution limit and it could be made smaller if needed, since the resolution of the measurement was set by the linewidth of the cw laser, to which the comb was locked. A scan over one interferogram lasted 2.5 s, and 150 steps were acquired, each of them averaged twice, with a total acquisition time of 20 min. The interleaved cavity spectra contain discrete spectral pieces, each covering 3 MHz and separated from each other by $f_{rep}^T = 1$ GHz.

The spectrum of an empty cavity spanning from 1500 to 1640 nm and containing 16000 resonance modes is shown in Fig. 2(a). The peak intensity of the cavity modes follows the comb envelope and a discrete peak is visible at $\lambda_{cw}$. The inset of Fig. 2(a) shows a zoom of a small part of the spectrum containing 3 cavity modes separated by $f_{rep}^{T}$. The spectrum of a cavity mode at 1600 nm is further enlarged in Fig. 2(b). A model based on a Lorentzian function and a linear baseline is fitted to the mode profile [red curve, with residuum in the lower panel]. The fitted parameters were the amplitude $A_q$, the center frequency $\nu_q$, and the width $\Gamma_q$ of the Lorentzian function, as well as the offset and slope of a linear baseline, and the fit returned $A_q$ = 1.019(3), $\nu_q$ = 187697701022.4(4) kHz, and $\Gamma_q$ = 214.4(9) kHz. To verify the accuracy of the mode width measurement, we measured the cavity ring-down time at 1600 nm directly after the cavity mode spectrum with the cavity unlocked from the cw laser, using the comb as a light source, an acousto-optic modulator before the cavity to interrupt the beam, and a monochromator with 200 GHz resolution to disperse the transmitted spectrum. The measurement yielded a mode width of $\Gamma_{RD}$ = 191(2) kHz, which agrees with the value obtained from the fit to the mode profile when considering the mode broadening due to the remaining cavity jitter (20 kHz) and the comb linewidth (2.8 kHz).

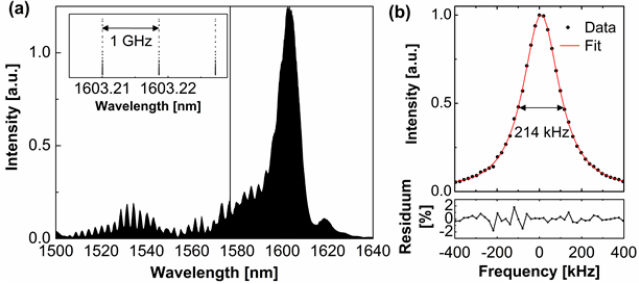

Fig. 2. (a) Spectrum of the empty cavity transmission spanning 16 THz. Inset: enlargement of 3 modes separated by 3FSR. (b) A zoom of a single mode at 1600 nm (black markers) together with a Lorentzian fit (red line) and the residuum (lower panel).

The center frequency of each of the 16000 cavity modes was determined from fits of a Lorentzian line shape function with an uncertainty ranging from 0.4 to 3 kHz depending on the signal-to-noise ratio of the considered mode. Figure 3(a) shows the shift of the cavity modes, calculated using Eqs (2) and (3), for the empty cavity [$\Delta\nu_0$, black curve, left y-axis in MHz] and the cavity filled with pure $N_2$ at 750(1) Torr [$\Delta\nu_{N2}$, red curve, right y-axis in GHz]. The reference FSR of an empty cavity, $FSR_{ref}^{0}$ in Eq. (2), was assumed to be equal to the $f_{rep}$ value (multiplied by 4/3) that maximized the transmission of the comb lines through the cavity modes around $\nu_{ref}= c/\lambda_{cw}$. This yielded $FSR_{ref}^{0}$ = 333.5730693(1) MHz and $f_0$ = 730.48(5) MHz. Note that these parameters are different for an empty cavity and cavity filled with $N_2$ because of the change of the intracavity refractive index, which had to be compensated for by adjusting the cavity length. This resulted in a change of $f_{rep}$ at $\nu_{ref}$ by $\Delta f_{rep}$ = -68.7900(1) kHz, and the reference FSR for the cavity filled with $N_2$ was calculated as $FSR_{ref}^{N_2}$ = $n_{ref}$ [4$\Delta f_{rep}$/3 + $FSR_{ref}^{0}$] = 333.5795270(2) MHz, where $n_{ref}$ is the refractive index of $N_2$ at $\nu_{ref}$, calculated using the Sellmeier equation with coefficients from [23]. The corresponding offset frequency was $f_0^{N2}$ = 552.98(5) MHz. Since both curves are referenced to $\nu_{ref}$, they cross 0 Hz at that frequency. The blue markers in Fig. 3(a) show the sum of the shift induced by $N_2$, calculated using Eq. (5) and the refractive index of $N_2$ from [23] [where the value $\nu_{ref}(n_{ref}-1)$ was subtracted to cross 0 Hz at $\nu_{ref}$], and the experimentally determined empty cavity shift, $\Delta\nu_0$. The excellent agreement of the calculated and the measured curves (relative error of the slopes of 4×10$^{-3}$) confirms the high accuracy of the measurement.

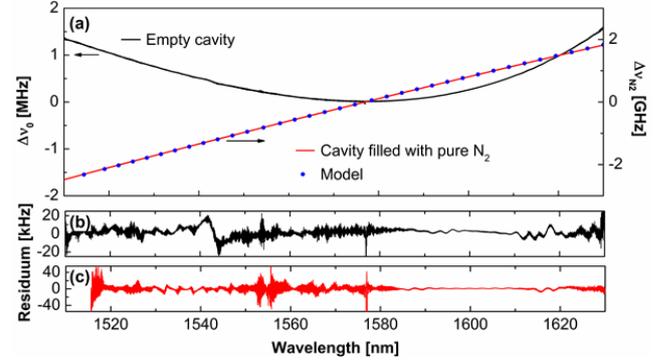

Fig. 3. (a) Shift of the cavity mode frequencies $\Delta\nu$ measured when the cavity is empty (black curve, left y-axis) and when the cavity is filled with pure $N_2$ at 750 Torr (red curve, right y-axis) plotted together with a calculated shift based on the Sellmeier equation for $N_2$ (blue markers). Note the three orders of magnitude difference between the two y-axis scales. (b) Residuum of a polynomial fit to the shift of the empty cavity modes. (c) Residuum of a polynomial fit to the mode shift of the cavity filled with $N_2$.

To retrieve the GDD, Eq. (3) was fitted to the two $\Delta\nu$ curves in Fig. 3 with $\Phi$ assumed as a fifth order polynomial function. The residuals of the fits are shown in Figs 3(b) and (c), confirming the validity of the model and the kHz precision of the determination of mode shift. The fitted curves were differentiated twice [see Eq. (4)] to yield the GDD of the empty cavity and cavity filled with $N_2$ shown in Fig. 4 by the black and red solid curves, respectively. The uncertainty of the GDD measurement is below 0.1 fs$^2$ over the entire range, originating from the uncertainty of the determination of $\nu_q$ and of the fit coefficients. The GDD of the cavity filled with $N_2$ is compared with a theoretical curve calculated as the sum of the experimentally determined empty cavity GDD and the GDD of $N_2$ calculated using the frequency-dependent refractive index of $N_2$ from [23] (dashed curve). Both curves agree within 1 fs$^2$.

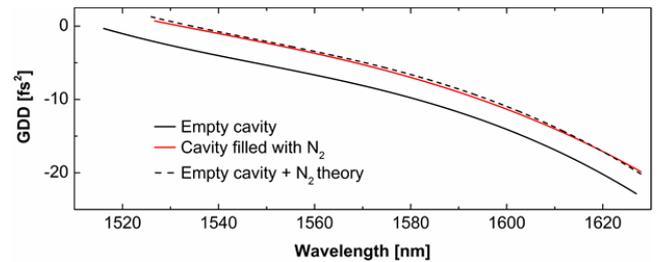

Fig. 4. Group delay dispersion (GDD) of the empty cavity (black solid curve), and the cavity filled with pure $N_2$ at 750 Torr (red solid curve). The dashed curve is the sum of the GDD of $N_2$ calculated using the Sellmeier and the empty cavity GDD (dashed black curve).

The high precision of the retrieved cavity mode frequencies allows also the measurement of dispersion induced by molecular transitions. Figure 5(a) shows the refractivity of the $3\nu_1+\nu_3$ band of $CO_2$ (black markers) obtained from the cavity mode frequencies measured when the cavity was filled with 1.00(1)% of $CO_2$ in $N_2$ at 750(1) Torr at room temperature [296(3) K] using Eq. (5). The red solid curve shows a fit of the molecular refractivity, Eq. (6), calculated using the imaginary part of the complex Voigt profiles and $CO_2$ line parameters from the HITRAN database [24], with the sample density as the fitting parameter. The non-resonant dispersion background was removed by fitting a fifth order polynomial function together with the model. The residual of the fit is shown in Fig. 5(b) and has a standard deviation equal to $5 \times 10^{-12}$ over the entire spectral range, demonstrating high precision of the measurement and a good agreement with theory. The $CO_2$ concentration retrieved from the fit was 0.95(1)%. The relative error of 5% with respect to the specified concentration, and the structure remaining in the residual, are presumably caused by the fact that the pressure broadening and shift parameters in HITRAN are defined for $CO_2$ in air while our sample was $CO_2$ in $N_2$.

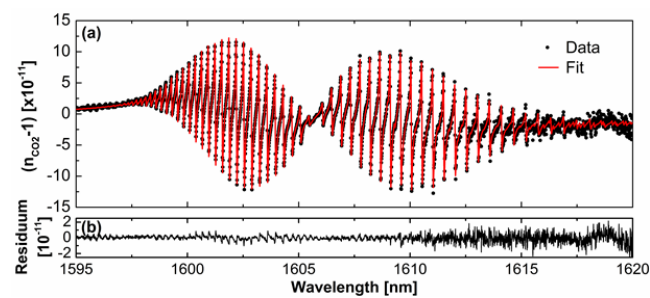

Fig. 5. (a) Refractivity of the $3\nu_1+\nu_3$ absorption band of $CO_2$ (black markers) together with a fit (red curve). (b) Residual of the fit.

To summarize, we measured cavity transmission spectra spanning 16 THz of bandwidth and containing 16000 modes with ~200 kHz linewidth using a comb-based Fourier transform spectrometer with sub-nominal resolution. Each cavity mode was fully characterized in terms of amplitude, width, and center frequency. From the shift of the center frequencies we retrieved the GDD of the cavity mirror coatings and of pure $N_2$, as well as the refractivity of the entire $CO_2$ absorption band. The precision of the GDD measurement was at a 0.1 $fs^2$ level and the accuracy of the $N_2$ measurement was within 1 $fs^2$ of a model based on the Sellmeier equation, while the precision of the refractive index of the molecular transitions was at the $5 \times 10^{-12}$ level.

Our method offers at least one order of magnitude improvement on the signal-to-noise ratio compared to the previous demonstration of direct measurement of cavity resonance modes using a dual-comb spectrometer [16], allowing retrieval of the cavity mode parameters with significantly improved precision. To our knowledge, our work demonstrates the narrowest spectral features ever measured with direct optical frequency comb spectroscopy, other than the comb lines themselves, which were measured using dual-comb spectroscopy [15]. It also proves that the resolution of a mechanical comb-based FTS is given by the comb linewidth, and not by the maximum optical path difference, provided that the nominal resolution of the FTS is matched to the repetition rate of the comb.

While the demonstrated precision of the dispersion measurement is comparable to that achieved with other comb-based methods of cavity dispersion determination [11-13], our approach offers at least 5 orders of magnitude improvement in frequency resolution. This in turn enables the measurement of cavity mode shifts induced by molecular transitions, which was previously possible only with cw laser-based techniques [5, 6]. The ability to measure dispersion of entire molecular bands without any prerequisite on the knowledge of the cavity parameters will allow determination of transition line parameters with improved accuracy.

In conclusion, direct measurement of cavity resonance modes using a comb-based Fourier transform spectrometer provides means to measure simultaneously the group delay dispersion of cavity mirror coatings, the dispersion of the refractive index of gases, and the resonant refractivity of entire molecular bands. Complementary information about the molecular transitions can be obtained from the linewidth and intensity of the cavity modes, opening up for full characterization of the real and imaginary parts of molecular bands and fundamental tests of the Kramers-Kronig relations.

**Funding.** Swedish Research Council (2016-03593) and (621-2015-04374), Swedish Foundation for Strategic Research (ICA12-0031), and the Knut and Alice Wallenberg Foundation (KAW 2015.0159).

**Acknowledgment.** GZ is supported by the China Scholarship Council.